\documentclass[aip,graphicx]{revtex4-1} 
\usepackage{graphicx}
\draft 
\usepackage{fancyhdr}
\begin{document}
\fancyhf{}
\renewcommand{\headrulewidth}{0pt}
\lfoot{\centering{~\thepage~}
	~\\
	~\\
	~\\
	Y.~Jin et al., Opt. Express \textbf{23}(16), 20418--20427 (2015)}

\title{Broadly, independent-tunable, dual-wavelength mid-infrared ultrafast 
	optical parametric oscillator}

\author{Yuwei Jin}
\author{Simona M. Cristescu}
\author{Frans J. M. Harren}
\author{Julien Mandon}
\email[]{j.mandon@science.ru.nl}
\homepage[]{\\ http://www.ru.nl/tracegasfacility/}

\affiliation{Molecular and Laser Physics, Institute for Molecules and Materials, Radboud University, P.O. Box 9010, 6500 GL Nijmegen, The Netherlands}

\date{\today}

\begin{abstract}
\label{abstract}
\textbf{Abstract}\\
We demonstrate a two-crystal mid-infrared dual-wavelength optical parametric 
oscillator, synchronously pumped by a high power femtosecond Yb:fiber laser. 
The singly-resonant ring cavity, containing two periodically poled lithium 
niobate crystals, is capable of generating two synchronized idler wavelengths, 
independently tunable over 30~THz in the 2.9 - 4.2~$\mu$m wavelength region, 
due to the cascaded quadratic nonlinear effect. The independent tunability of 
the two idlers makes the optical parametric oscillator a promising source for 
ultrafast pulse generation towards the THz wavelength region, based on 
different frequency generation. In addition, the observed frequency doubled 
idler within the crystal indicates the possibility to realize a broadband 
optical self-phase locking between pump, signal, idler and higher order 
generated parametric lights. 	
\end{abstract}

\maketitle 

\section{Introduction}
\label{section_1}      
Dual-wavelength operation within a single laser cavity is promising for a 
variety of applications such as coherent pulse synthesis, THz generation, 
pump-probe experiments and CARS\cite{Ma2001,Kawase2000,Yu2005,Ganikhanov2006}. 
For conventional solid-state ultrafast lasers, dual-wavelength operations were 
observed and demonstrated with Ti:sapphire lasers at 
800~nm\cite{Barros1993,Zhang1993} and Tm:CaYAlO$_4$ lasers at 
2~$\mu$m\cite{Kong2015}. Besides, dual-wavelength operation was also realized 
with a $Q$-switched cascade fiber laser at 3~$\mu$m\cite{Li2012}. Optical 
parametric oscillators (OPOs) are well-established mid-infrared sources 
offering coherent light, high power, broad bandwidth and broad tuning range. 
Nowadays, OPOs are also widely used for a variety of applications such as 
mid-infrared frequency comb generation and 
spectroscopy\cite{Adler2009,Foltynowicz2011,Jin2014}. Specifically, 
synchronously pumped OPOs offer an new opportunity for synchronized pulses 
generation at two different wavelength 
regions\cite{Burr1997,Sun2006,Tartara2007,Samanta2011,Xu2012,Jiang2013,
	Ramaiah-Badarla2014,Gu2014}, that can be directly used for THz generation based 
on different frequency generation 
(DFG)\cite{Taniuchi2004,Schaar2007,Vodopyanov2008,Vodopyanov2011,Hegenbarth2012}
. In the case of an OPO with a single cavity, the dual-wavelength operation is 
realized when the two resonant pulse trains have the same round trip time 
delay\cite{Burr1997}. In literature, the simultaneous dual-wavelength operation 
is explained by the balance between phase matching and group-velocity 
mismatching between the two resonant pulses\cite{Tartara2007,Xu2012}. 
Experimental results also indicate a stable relative carrier-envelope 
phase-slip frequencies within a dual-wavelength femtosecond OPO, which is of 
great interest for various applications such as quantum control and coherent 
pulse synthesis\cite{Sun2006}. However, most dual-wavelength OPOs reported 
provide limited dual-wavelength tunability and broadband arbitrary wavelength 
tuning of the two idelers is not possible. 

In this paper, we present the experimental observation of dual-wavelength 
operation within a two-crystal OPO pumped by a femtosecond Yb:fiber laser. By 
tuning the cavity length, the simultaneous dual-wavelength operation is 
realized when the two crystals are at different poling periods. To our 
knowledge, this is the first demonstration of an independently-tunable, 
synchronously-pumped, dual-wavelength, ultrafast OPO generating arbitrarily 
tunable idlers across a 30~THz spectral region between 2.9~$\mu$m and 
4.2~$\mu$m. To characterize the OPO, we have also observed the behavior of 
several cascaded parametric processes within the nonlinear crystals, such as 
the second harmonic generation(SHG) of the idler, indicating the possibility to 
achieve optical self-phase locking between pump, signal and idler when the 
spectrum of the frequency doubled idler overlaps the spectrum of the signal. It 
is worth mentioning that previously, self-phase locking of continuous-wave (CW) 
OPOs has been achieved by using the intra-cavity frequency-doubled 
idler\cite{Lee1999,Zondy2001,Douillet2001,Lee2003}. 

\section{Dual-wavelength operation of the optical parametric oscillator}
\label{section_2}  
\begin{figure}[!b]
	\centering\includegraphics[width=12cm]{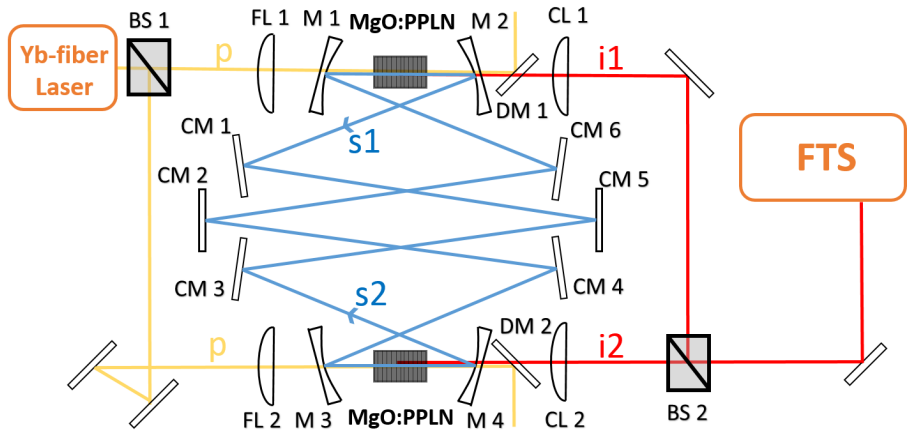}
	\caption{Scheme of the femtosecond laser pumped dual-wavelength OPO. p, 
		s1-s2, i1-i2, pump, signals, and idlers, respectively. BS1-BS2, beam splitters. 
		FL1-FL2, focusing lenses (f = 100~mm). M1-M4,curved mirrors (R = 100~mm). 
		CM1-CM6, chirped mirrors. DM1-DM2, dichroic mirrors. CL1-CL2, collimating 
		lenses (f = 100~mm). FTS, Fourier transform spectrometer. One moving stage is 
		mounted on CM 5 (not shown).}
	\label{fig_1}
\end{figure} 
\begin{figure}[!b]
	\centering\includegraphics[width=12cm]{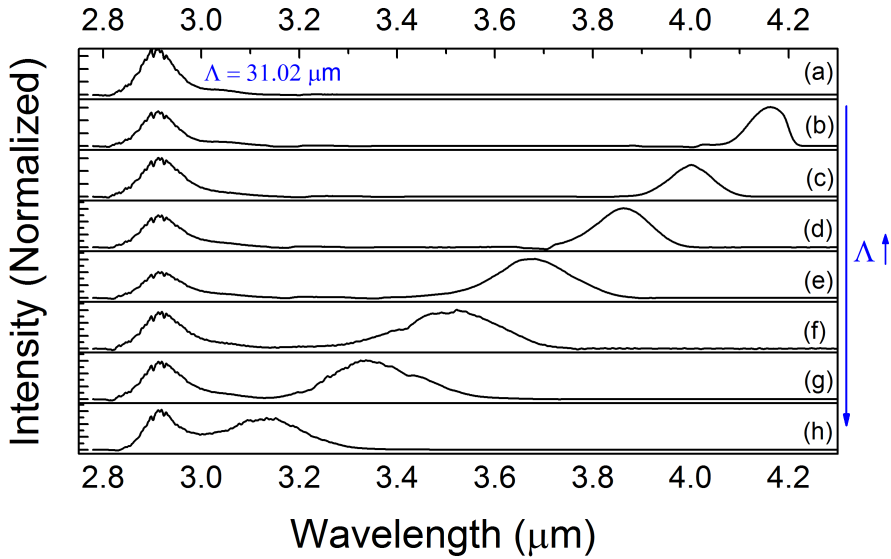}
	\caption{Panel (a): Generated idler light with one crystal pumped. 
		Panel (b)-(h): Generated idler light with two crystals pumped, in which the 
		poling period of the second crystal is varied between 27.91 - 30.49 ~$\mu$m. 
		Spectra are measured by the FTS.}
	\label{fig_2}
\end{figure}

\noindent The experimental setup of the synchronously pumped singly-resonant 
OPO is similar to the one described previously \cite{Jin2014,Jin2015}. As 
illustrated in Fig. \ref{fig_1}, the difference is that the two crystals (both 
temperatures are stabilized at 80~degree) are pumped by a single 
femtosecond Yb-fiber laser instead of two similar pump lasers in order to have 
exactly the same repetition rates of the pump pulses for the crystals. 
The two generated idlers are recombined together with a mid-infrared beam 
splitter (BS 2), coupling to a commercial Fourier transform spectrometer (FTS) 
for characterization. The two resonating signals are counter-propagating in the 
ring cavity and do not interact with one another  within the crystals. The zero 
dispersion wavelength of the cavity is calculated to be at 1532~nm (195.7~THz), considering 
both the crystal dispersion\cite{Jundt1997} and the six chirped mirrors.

To measure the output spectra of the OPO, the poling period of one crystal is 
fixed at 31.02~$\mu$m, while the poling period of the second crystal is tuned 
from 27.91~$\mu$m to 30.49~$\mu$m. Figure 2(a) shows the measured spectrum when 
only one crystal with a 31.02~$\mu$m poling period is directly pumped by the 
femtosecond laser. When the two crystals are pumped by the femtosecond laser, 
two idlers are generated simultaneously within the two crystals at different 
wavelengths. It can be noticed that the first idler is not spectraly affected 
when both crystals are pumped. By changing the poling period of the second 
crystal, dual-wavelength operation across a 30~THz spectral region between 2.9 
- 4.2~$\mu$m is realized (Fig.~\ref{fig_2}(b) - \ref{fig_2}(h)). The two idler wavelengths 
are operating independently and as such can be tuned to arbitrary wavelength 
regions.
\begin{figure}[!b]
	\centering\includegraphics[width=12cm]{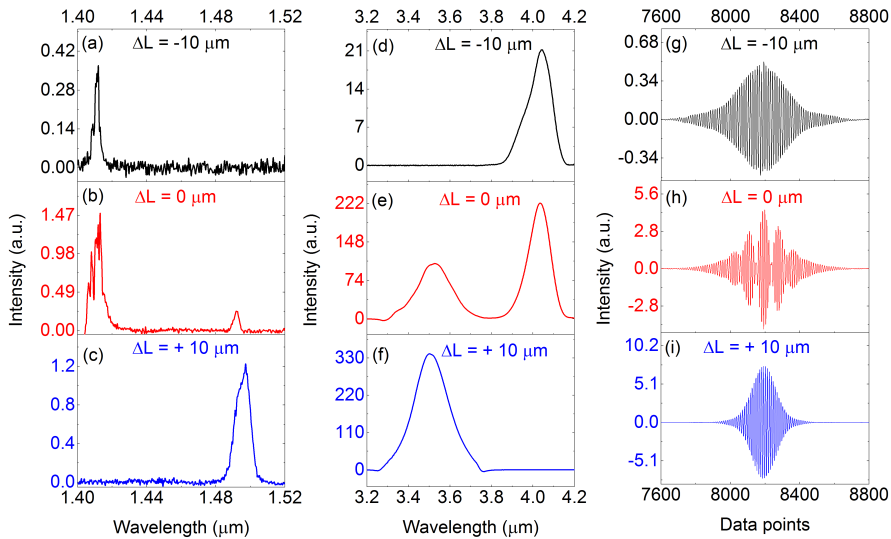}
	\caption{Panel (a)-(c): Measured signal spectra by tuning the cavity 
		length. Panel (d)-(f): Measured idler spectra by tuning the cavity length. 
		Panel (g)-(i): Corresponding interferograms of idlers retrieved from the FTS by 
		tuning the cavity length.}
	\label{fig_3}	
\end{figure}
 
The behavior of the dual-wavelength operation is cavity length dependent. 
Measurements have been done when the poling periods of the two crystals are set 
to be 28.28 and 29.52~$\mu$m. As illustrated in Fig.~\ref{fig_3}, by tuning the 
cavity length, both the single-wavelength and the dual-wavelength operations 
are achieved when the pump power is 1.1~W. Figures~\ref{fig_3}(b), \ref{fig_3}(e) and \ref{fig_3}(h) exhibit 
the signal spectrum, the idler spectrum and the interferogram of the idler 
retrieved from the FTS, respectively, for which a relative cavity length 
$\Delta L = 0~\mu m$ is assumed (i.e. the cavity length that is synchronized to 
the repetition frequency of the pump laser). Moreover, each OPO could deliver 
pulses singly at either wavelength by detuning the cavity length as can be seen 
in Fig.~\ref{fig_3}(a), \ref{fig_3}(d), \ref{fig_3}(g) and Fig.~\ref{fig_3}(c), \ref{fig_3}(f), \ref{fig_3}(i). The reason the 
relative intensities of the signals are quite weak is the usage of the 
high-reflectivity mirrors of the OPO cavity.

\section{Characterization of the optical parametric oscillator}
\label{section_3}      

\begin{figure}[!b]
	\centering\includegraphics[width=12cm]{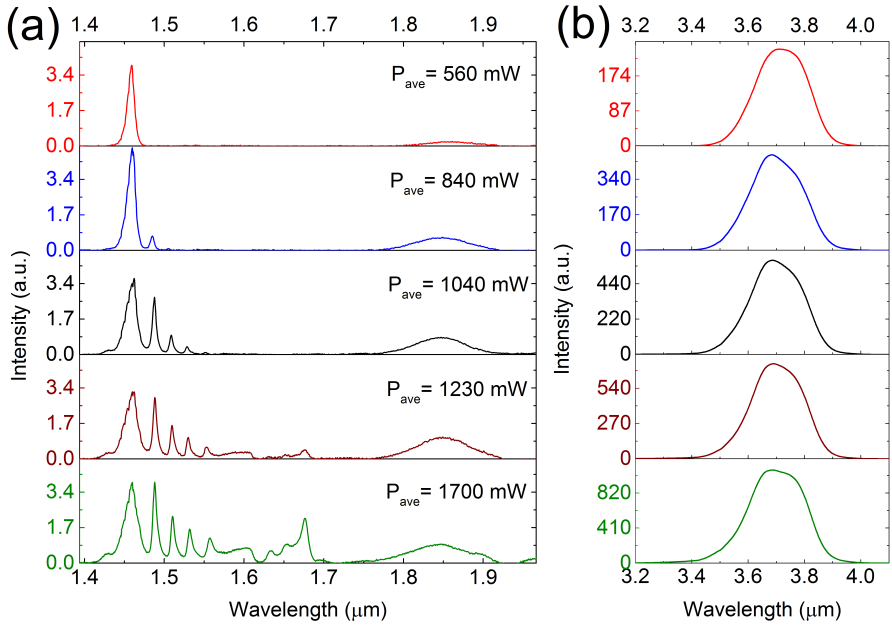}
	\caption{Panel (a):Recorded spectra between 1.3 and 2.0~$\mu$m under 
		different pump powers. The peaks between 1.8 and 1.9~$\mu$m are the frequency 
		doubled idlers. The lights between 1.4 and 1.7~$\mu$m are recorded signals. 
		Panel (b): The corresponding idler spectra between 3.2 and 4.1~$\mu$m under 
		different pump powers.}
	\label{fig_4}	
\end{figure}

\noindent It is worth noticing that the arbitrary wavelength tuning of a 
dual-wavelength OPO within a single cavity cannot be explained from the 
conventional perspective\cite{Burr1997,Tartara2007,Xu2012}, hence, more 
experiments have been done to have an in-depth characterization of the OPO. As 
the two resonating signals are counter-propagating in the ring cavity and do 
not interact with one another, the beam splitter (BS 1 in Fig. \ref{fig_1}) can 
be removed to characterize the two OPOs separately. Therefore, only one crystal 
is pumped directly by the Yb-fiber laser. From the output of the OPO, idler in 
the mid-infrared, signal and frequency doubled idler in the near infrared can 
be observed simultaneously by using a FTS. In addition, other cascaded 
parametric lights below 1 $\mu$m can be monitored by a visible light 
spectrometer. Power scaling measurements are performed when the poling periods 
of the two crystals are set to be 29.08~$\mu$m. As illustrated in Fig.~
\ref{fig_4}(a), the spectra of signals between 1.4 and 1.7~$\mu$m and the 
spectra of frequency doubled idlers between 1.8 and 1.9~$\mu$m are recorded.  
The corresponding idler spectra between 3.2 and 4.1~$\mu$m are also shown in 
Fig. \ref{fig_4}(b). A spectral broadening effect is observed while increasing 
the pump power. At a 1.7~W average pump power, the signal covers a 300~nm 
spectral range corresponding to a 1000~cm$^{-1}$ (30~THz) bandwidth, revealing a 
modulated multi-peak structure. Both the spectral broadening effect and the 
modulated multi-peak structure indicate the presence of a nonlinear phase 
modulation, which is plausibly due to the cascaded quadratic nonlinearity for 
the signal that is treated as the fundamental wave in this case\cite{Zhang2015}. 
These cascaded quadratic processes have been shown to resemble a $\chi^{(3)}$ 
process\cite{Menyuk1994,Torner1996,Iizuka1999,Desalvo1992,Boyd2008}, and the 
most common cascaded quadratic effect involves a single-step process, such as 
non-phase-matched SHG in a dispersive quadratic media. Meanwhile, cascaded 
quadratic processes due to third-harmonic generation(THG) have also been 
discussed previously\cite{Koynov1998,Saltiel1999,Saltiel2000}. 

\begin{figure}[!b]
	\centering\includegraphics[width=12cm]{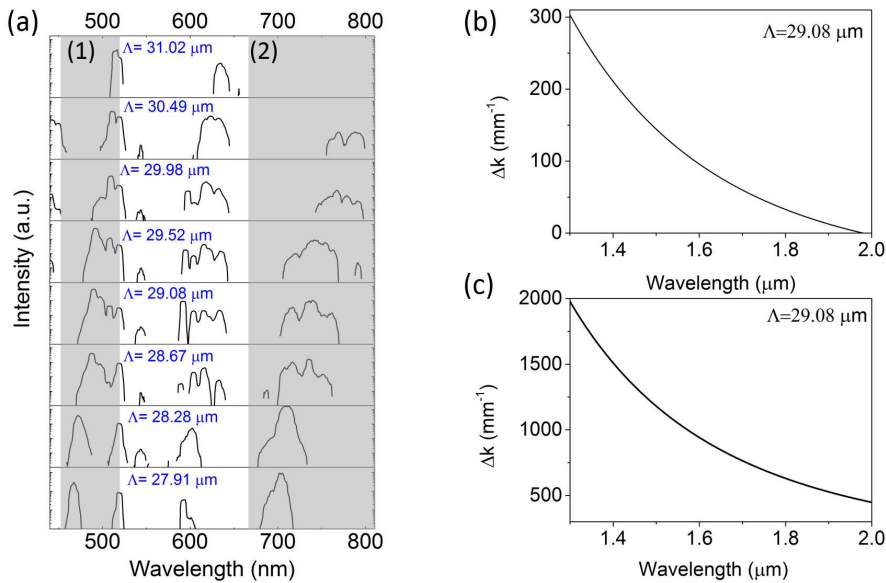}
	\caption{Panel (a): Recorded spectra between 400 and 800~nm for different 
		poling periods of the PPLN crystal. All vertical axes are at logarithmic 
		scales. Panel (b): Phase mismatch of SHG when the fundamental wave is at 
		different wavelength between 1.3 and 2.0~$\mu$m. Panel (c): Phase mismatch of 
		THG when the fundamental wave is at different wavelength between 1.3 and 
		2.0~$\mu$m.}
	\label{fig_5}
\end{figure}

To further confirm the presence of the cascaded quadratic processes, a series 
of measurements are performed with a 1.7~W pump power. By tuning the poling 
periods of two crystals from 27.91 to 31.02~$\mu$m, spectra between 400 and 
800~nm (375 to 749~THz) are recorded. In these measurements, both the THG and SHG of the signal 
are observed as demonstrated in gray sections (1) and (2) within Fig.~
\ref{fig_5}(a), respectively. In this case, both the phase-mismatching for SHG 
and THG of the signal can be calculated by the following formulae 
$\Delta~k_1=2\pi(\lambda_{2s}/n_{2s}-2\lambda_{s}/n_{s}-1/\Lambda)$ and 
$\Delta~k_2=2\pi(\lambda_{3s}/n_{3s}-3\lambda_{s}/n_{s}-1/\Lambda)$, 
respectively, where $\lambda_{s}$, $\lambda_{2s}$, $\lambda_{3s}$ are the 
wavelengths of the signal, the SHG of the signal, and the THG of the signal; 
$n_{s}$, $n_{2s}$, $n_{3s}$ are corresponding refractive indices; $\Lambda$ 
represents poling period of the crystal. Figure~\ref{fig_5}(b) and 
\ref{fig_5}(c) give an example of the results upon these calculations, assuming 
that the poling period is 29.08~$\mu$m. The observation of both 
non-phase-matched SHG and THG of the resonating signal implies the presence of 
cascaded quadratic processes which we believe contribute to the dual-wavelength 
operation.  

\begin{figure}[!b]
	\centering\includegraphics[width=12cm]{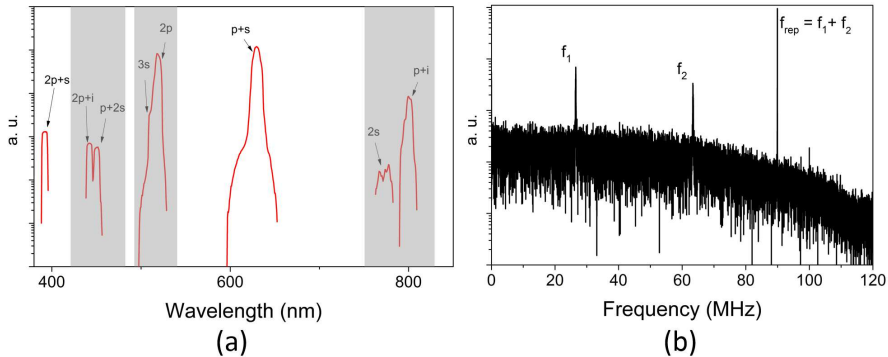}
	\caption{Panel (a): Recorded spectrum of the cascaded parametric light 
		generated from the OPO along with the residual pump(p), leaked signal(s) and 
		generated idler(i). Within the grey sections,there are overlaps between the 
		different cascaded parametric lights, at some poling periods. Panel (b): 
		Beating signal (log scale) detected by using a single visible light detector, 
		$f_1 = f_p-3f_i$ or $f_2 = f_p-3f_i$. All vertical axes are at logarithmic 
		scales.}
	\label{fig_6}
\end{figure} 

When the poling period of the pumped crystal is 30.49~$\mu$m with a 300~mW 
average pump power, a spectrum of the cascaded parametric lights generated from 
the OPO along with the residual pump light ($p$), leaked signal light ($s$) 
from the cavity, and the generated idler light ($i$), are recorded with an 
optical spectrum analyzer (OSA). As illustrated in Fig. \ref{fig_6}(a), the 
dominating, nonlinearly generated intensities are: the sum frequencies of pump 
and signal (p+s), pump and idler, and doubling frequency of pump ($2p$), next 
to other frequencies ($2s$, $3s$, $p+2s$, $2p+i$, $2p+s$).
By monitoring the light leaking from mirror CM3 by using a fast visible light 
detector, two beating frequencies are observed centered on half the repetition 
frequency (45~MHz) of the pump laser (Fig. \ref{fig_6}(b)). This beating 
frequency originates from the overlap between different generated parametric 
lights as shown in the gray sections in Fig. \ref{fig_6}(a). Assume that the 
frequencies of the pump and the idler are $f_p$ and $f_i$, the beating 
frequencies observed at the three gray sections are $f_{2p+i} - f_{p+2s}$, 
$f_{3s} - f_{2p}$, and $f_{2s} - f_{p+i}$, respectively. The value of these 
beating frequencies are all equal to $f_p - 3f_i$, considering that $f_{p}=f_{s} 
+ f_{i}$. Moreover, the same beating frequency in the mid-infrared wavelength 
region is also observed by using a fast infrared detector (Vigo, PVI-4TE, 50~MHz 
bandwidth) at the output of the OPO after a Germanium collimating lens.

\begin{figure}[!b]
	\centering\includegraphics[width=11cm]{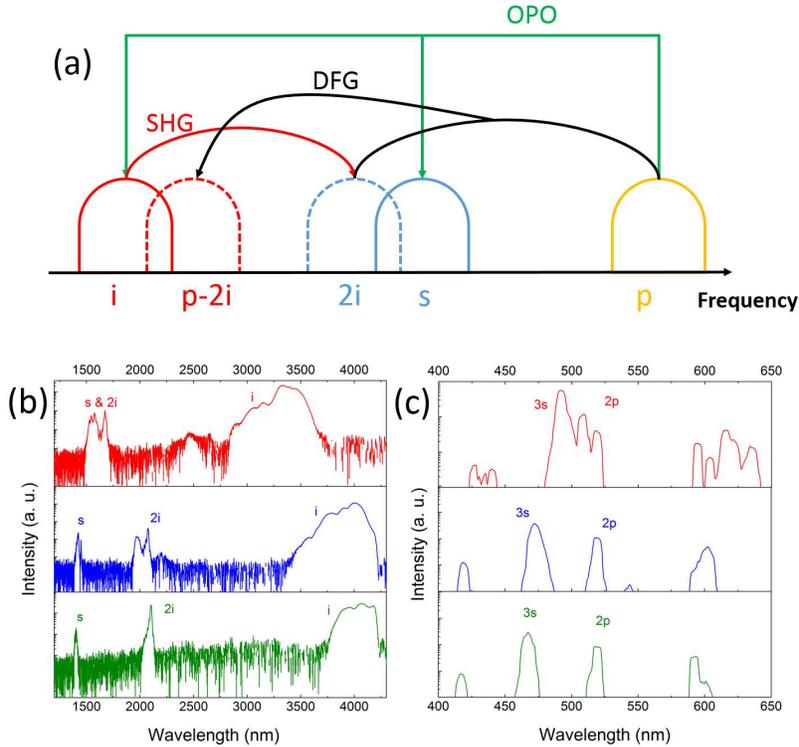}
	\caption{Panel (a): Parametric processes within the OPO cavity. OPO: 
		optical parametric oscillation; DFG: different frequency generation; SHG: 
		second harmonic generation. p,s,and i: pump, signal and idler. Panel (b): 
		Spectra measured between 1000~nm and 5000~nm (60 to 300~THz) at three different poling periods, 
		indicating the merging between the signal ($s$) and the second harmonic of 
		idler ($2i$). Panel (c): Spectra measured between 400~nm and 650~nm (461 to 749~THz) at the 
		three different poling periods, indicating the merging between the third 
		harmonic of signal ($3s$) and the second harmonic of pump ($2p$). All vertical 
		axes are at logarithmic scales.}
	\label{fig_7}
\end{figure}

To explain the observed beating frequencies both in the mid-infrared and 
visible regions, the parametric processes within the OPO are illustrated in 
Fig.~\ref{fig_7}(a). Firstly, the pump wavelength ($f_p$) is divided into a 
signal wavelength ($f_s$) and an idler wavelength ($f_i$). The SHG of the idler 
is at $2f_i$ while the different frequency generation (DFG) between the 
frequency doubled idler ($2f_i$) and the pump ($f_p$) is at $f_p-2f_i$. By 
changing the poling periods in the crystal at 27.91~$\mu$m, 28.28~$\mu$m and 
29.52~$\mu$m, the frequency doubled idler ($2f_i$) and the signal ($f_s$) start 
to merge as can be seen in Fig.~\ref{fig_7}(b), generating a beating frequency 
at $f_p - 3f_i$. Figure~\ref{fig_7}(c) illustrates the merging of the THG of the 
signal ($3f_s$) with the SHG of the pump (2$f_p$), also generating a beating 
frequency at $f_p - 3f_i$. In addition, the beating frequency detected in the 
mid-infrared is calculated to be $f_{p-2i} - f_i = f_p - 2f_i- f_i =f_p - 
3f_i$, which is the same value when compared to the beating frequencies both in 
the near infrared and the visible wavelength region. It is expected that from 
the overlap between the signal and the frequency doubled idler 
(Fig.~\ref{fig_7}(a)), self-phase locking can be achieved from this femtosecond 
laser pumped OPO, and that all generated parametric lights can be optically 
phase locked to the pump source. One way to establish this is by stabilizing the 
cavity length with a feedback loop \cite{Wong2008}. Once the self-phase locking 
is realized, $f_{p}$ is constantly equal to $3f_i$, and one can achieve phase 
coherent generation at frequencies of $N\cdot f_i (N = 1 - 8)$  from a single 
OPO laser source, simultaneously, which are illustrated in Table 1. 
\begin{table}[!t]
	\centering\caption{The corresponding frequencies of different cascaded 
		parametric lights when the OPO would be self-phase locked between the frequency 
		doubled idler and the signal. $f_i$ is the idler frequency. }
	\begin{tabular}{lp{1.2in}|lp{1.2in}}
		\hline
		Parametric lights        & Frequencies                & 
		Parametric lights        & Frequencies          \\ \hline
		$i$ and $p-2i$   &  $f_i$              & $p+s$  &  $5f_i$     \\
		$s$ and $2i$  &     $2f_i$       & $3s$ and $2p$  &  $6f_i$     
		\\
		$p$      &    $3f_i$          & $2p+i$ and $p+2s$  &  $7f_i$    
		\\
		$2s$ and $p+i$ &    $4f_i$    & $2p+s$   &   $8f_i$      \\ 
		\hline
	\end{tabular}
\end{table}

\section{Discussion and conclusion}
\label{section_4}      
For a femtosecond OPO with a single gain medium, pumped at a fixed repetition 
frequency, the central wavelength of the signal wave shifts by changing the 
cavity length, maintaining the round trip time delay corresponding to the fixed 
cavity length\cite{Tartara2007,Jin2015}. Asynchronous pumping can also be 
achieved at a fixed cavity length by using two femtosecond lasers with 
different repetition frequencies\cite{Jin2015,Zhang2012}. In this case, two 
pulse trains (signal) at different central wavelengths maintain the same round 
trip time delay due to the difference of group velocities within the crystal. 
Conventionally, dual-wavelength operation within a single cavity at one 
repetition frequency can be realized when the two resonant pulse trains have 
the same round trip time delay, implying that the tuning of the two wavelengths 
is strongly correlated and not arbitrary\cite{Burr1997,Tartara2007,Xu2012}. In 
this work, the two-crystal OPO offers arbitrary dual-wavelength generation, due 
to the use of different poling periods in the crystals offering different gain 
bandwidths. Next to this, the OPO is singly-resonant on the signal without 
output coupler, reserving extremely high signal power within the cavity. This 
two crystal OPO cavity is capable of generating two synchronized idler beams 
that are independently tunable across a 30 THz spectral region between 2.9 - 
4.2~$\mu$m. The independent tunability of the two idlers can be explained by 
taking higher order nonlinear effects into account, such as the cascaded 
quadratic nonlinearity. These cascaded processes in quadratic media induce a 
nonlinear phase shift on the fundamental wave by back-conversion of the 
phase-mismatched second harmonic wave, which is equivalent to third-order 
nonlinearity\cite{Desalvo1992,Boyd2008,Buryak2002,Marangoni2006}. In practice, 
active group velocity control within a PPLN crystal has already been achieved 
based on the cascaded quadratic nonlinearity\cite{Marangoni2006,Lu2008}. We 
believe that the group velocity mismatch between the two wavelengths are 
compensated by the nonlinear phase shift, due to the cascaded quadratic 
nonlinearities that is analogous to the third-order nonlinearity in a 
conventional mode-locked laser in the case of Kerr 
modulation\cite{Boyd2008,Diels2006}.

In conclusion, for the first time, we have experimentally demonstrated 
dual-wavelength operation within a two-crystal OPO cavity, generating two 
broadband idler beams that are independently tunable across a 30~THz spectral 
range between 2.9 - 4.2~$\mu$m. Such passively synchronized dual-wavelength 
pulses could be coupled to a nonlinear crystal, such as 
OP-GaAs\cite{Vodopyanov2008}, to generate light from the far-infrared to THz 
wavelength region based on the different frequency generation. Besides, the 
spatially separated pulses are also ideal sources for mid-infrared pump-probe 
experiments. To characterize the OPO, we have observed second harmonic 
generation of the idler ($2i$) directly, without using a tandem crystal. The 
observed merging effect between the signal and the frequency doubled idler 
indicates a possibility to realize self-phase locking between them. 

\section*{Acknowledgments}
\label{section_5} 
\noindent This work is financially supported by the Technology Foundation STW, under 
project number 11830.

\end{document}